\documentclass[aps,amsmath,twocolumn,floatfix,showpacs,superscriptaddress,noshowpacs]{revtex4}
\usepackage{amsfonts}
\usepackage{amsmath}
\usepackage{amsthm}
\usepackage{amssymb}
\usepackage{graphicx}
\usepackage[T1]{fontenc}
\usepackage{supertabular}
\usepackage{longtable}
\usepackage[usenames,dvipsnames]{color}
\usepackage{bbm}
\usepackage{fancyhdr}
\usepackage{breqn}
\usepackage{fixltx2e}
\usepackage{capt-of}
\usepackage{tikz}
\usetikzlibrary{matrix}
\usepackage{endnotes}
\usepackage{soul}
\usepackage{marginnote}

\newcommand{\mathsym}[1]{}
\newcommand{\unicode}[1]{}

\usepackage{pdfpages}
\usepackage{enumitem}
\usepackage[sort&compress]{natbib}
\usepackage{graphicx}

\usepackage{float}

\usepackage{xcolor}
 
\usepackage[framemethod=TikZ]{mdframed}
\usepackage[framemethod=TikZ]{mdframed}
\usepackage{framed}
\mdfsetup{%
	outerlinewidth=1,skipabove=20pt,backgroundcolor=yellow!50, outerlinecolor=black,innertopmargin=0pt,splittopskip=\topskip,skipbelow=\baselineskip, skipabove=\baselineskip,ntheorem,roundcorner=5pt}

\mdtheorem[nobreak=true,outerlinewidth=1,%leftmargin=40,rightmargin=40,
backgroundcolor=yellow!50, outerlinecolor=black,innertopmargin=0pt,splittopskip=\topskip,skipbelow=\baselineskip, skipabove=\baselineskip,ntheorem,roundcorner=5pt,font=\itshape]{result}{Result}

\mdtheorem[nobreak=true,outerlinewidth=1,%leftmargin=40,rightmargin=40,
backgroundcolor=yellow!50, outerlinecolor=black,innertopmargin=0pt,splittopskip=\topskip,skipbelow=\baselineskip, skipabove=\baselineskip,ntheorem,roundcorner=5pt,font=\itshape]{theorem}{Theorem}

\mdtheorem[nobreak=true,outerlinewidth=1,%leftmargin=40,rightmargin=40,
backgroundcolor=gray!10, outerlinecolor=black,innertopmargin=0pt,splittopskip=\topskip,skipbelow=\baselineskip, skipabove=\baselineskip,ntheorem,roundcorner=5pt,font=\itshape]{remark}{Remark}

\mdtheorem[nobreak=true,outerlinewidth=1,%leftmargin=40,rightmargin=40,
backgroundcolor=pink!30, outerlinecolor=black,innertopmargin=0pt,splittopskip=\topskip,skipbelow=\baselineskip, skipabove=\baselineskip,ntheorem,roundcorner=5pt,font=\itshape]{quaestio}{Quaestio}

\mdtheorem[nobreak=true,outerlinewidth=1,%leftmargin=40,rightmargin=40,
backgroundcolor=yellow!50, outerlinecolor=black,innertopmargin=5pt,splittopskip=\topskip,skipbelow=\baselineskip, skipabove=\baselineskip,ntheorem,roundcorner=5pt,font=\itshape]{background}{Background}

\begin{document}

\title{Combining PCR and CT testing for COVID}

\author{Chen Shen}
\affiliation{New England Complex Systems Institute}
\author{Ron Mark MD  DABR}
\affiliation{Mark Medical Care PLLC}
\author{Nolan J. Kagetsu MD DABR}
\affiliation{Mt. Sinai Hospital}
\author{Anton S. Becker MD PhD}
\noaffiliation
\author{Yaneer Bar-Yam}
%\email[E-mail: ]{yaneer@necsi.edu}
\affiliation{New England Complex Systems Institute}

\date{May 26, 2020}

\begin{abstract}
We analyze the effect of using a screening CT-scan for evaluation of potential COVID-19 infections in order to isolate and perform contact tracing based upon a viral pneumonia diagnosis. RT-PCR is then used for continued isolation based upon a COVID diagnosis. Both the low false negative rates and rapid results of CT-scans lead to dramatically reduced transmission. The reduction in cases after 60 days with widespread use of CT-scan screening compared to PCR by itself is as high as $50\times$, and the reduction of effective reproduction rate $R(t)$ is $0.20$.  Our results imply that much more rapid extinction of COVID is possible by combining social distancing with CT-scans and contact tracing.
\end{abstract}

\maketitle

Testing and isolation is a foundation of COVID-19 prevention. However, the most common test, PCR, has a false negative rate often reported to be $30\%$ \cite{CT, PCR-1,PCR-2}, varying with disease period, sampling method and processing materials. Individuals who test negative may resume normal activities with family, roommates or in essential services. Moreover, PCR test results typically take several days. Rapid point of care systems, available in some locations, use the same sampling and analysis methods so false negative rates are likely similar or worse. Recent reports indicate $48\%$ \cite{abbott}. Serological tests are generally not positive at onset of symptoms and do not indicate that an individual is infectious at the time of the test. Since effective at home isolation from family and roommates is difficult, isolation that is effective may not be performed prior to receiving test results. Both the time delay and those who are false negative may lead to many additional new cases. Mild cases that do not progress to severe may persist in being infectious for extended periods of time. These individuals are highly contagious and therefore more harmful to transmission control than asymptomatic carriers. 

CT-scans have a reported small false negative rate for viral pneumonia, including COVID, at onset of fever or respiratory symptoms \cite{initial}. Test results are available within minutes, and CT equipment is widespread \cite{cteq}. For example, there are about 15,000 CT imaging devices in the US, of which over 5,000 are in ambulatory care locations. The scan itself takes seconds to perform, readings in minutes. The time needed for cleaning and disinfecting \cite{ACR} can be mitigated (Appendix). Efficient intake can enable hundreds of tests per device per day \cite{speed}. The potential capacity of millions is large compared to the current PCR testing of under 400,000 per day \cite{PCRtests}. CT devices and radiological services are underutilized currently due to deferral of usual medical care. With a positive rate for widely available testing only a few percent \cite{Germany,SouthKorea,Sicily}, conclusive test results promote focusing on ensuring those who do test positive are isolated. 

Figures 1 and 2 compare the impact of using CT-scans and PCR on disease spread based upon a standard model of transmission and social isolation for moderate and strong social distancing. The daily infection rate is shown for (red) PCR tests of only severe cases, (orange) PCR tests for all cases regardless of severity, (blue) combined use of CT-scans (as initial screening) and PCR. Dotted lines include contact tracing quarantining $50\%$ of close contacts.

\begin{figure}[t]
\centering
\includegraphics[scale=0.20]{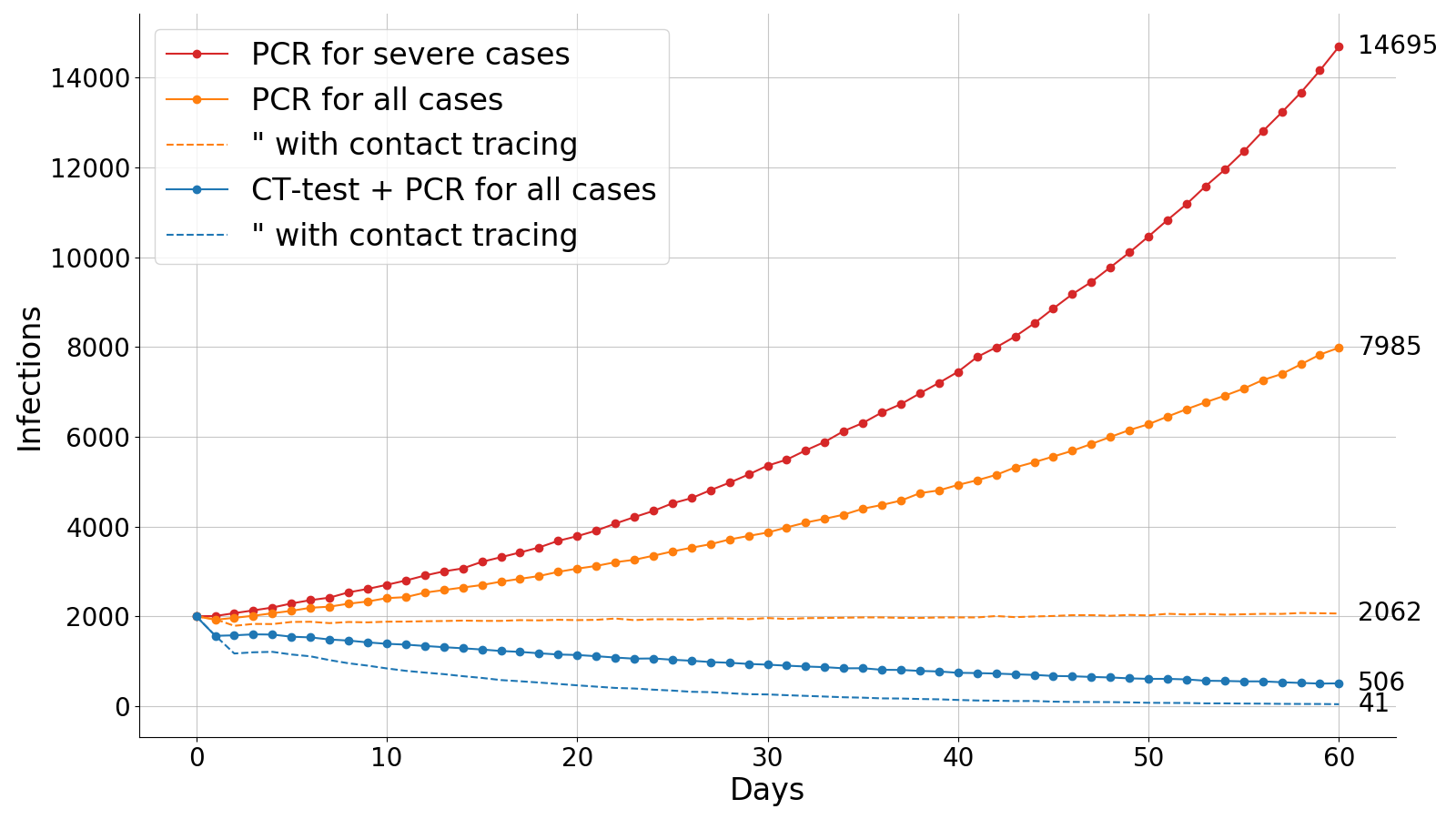}
\caption{Daily new infections with three testing strategies and without and with contact tracing, average of 60 simulation runs (see text). For dotted lines $50\%$ of close contacts are quarantined. Reference reproduction rate due to moderate social interventions is $R^*=1.25$ (see methods). }
\end{figure}

\begin{figure}[t]
\centering
\includegraphics[scale=0.20]{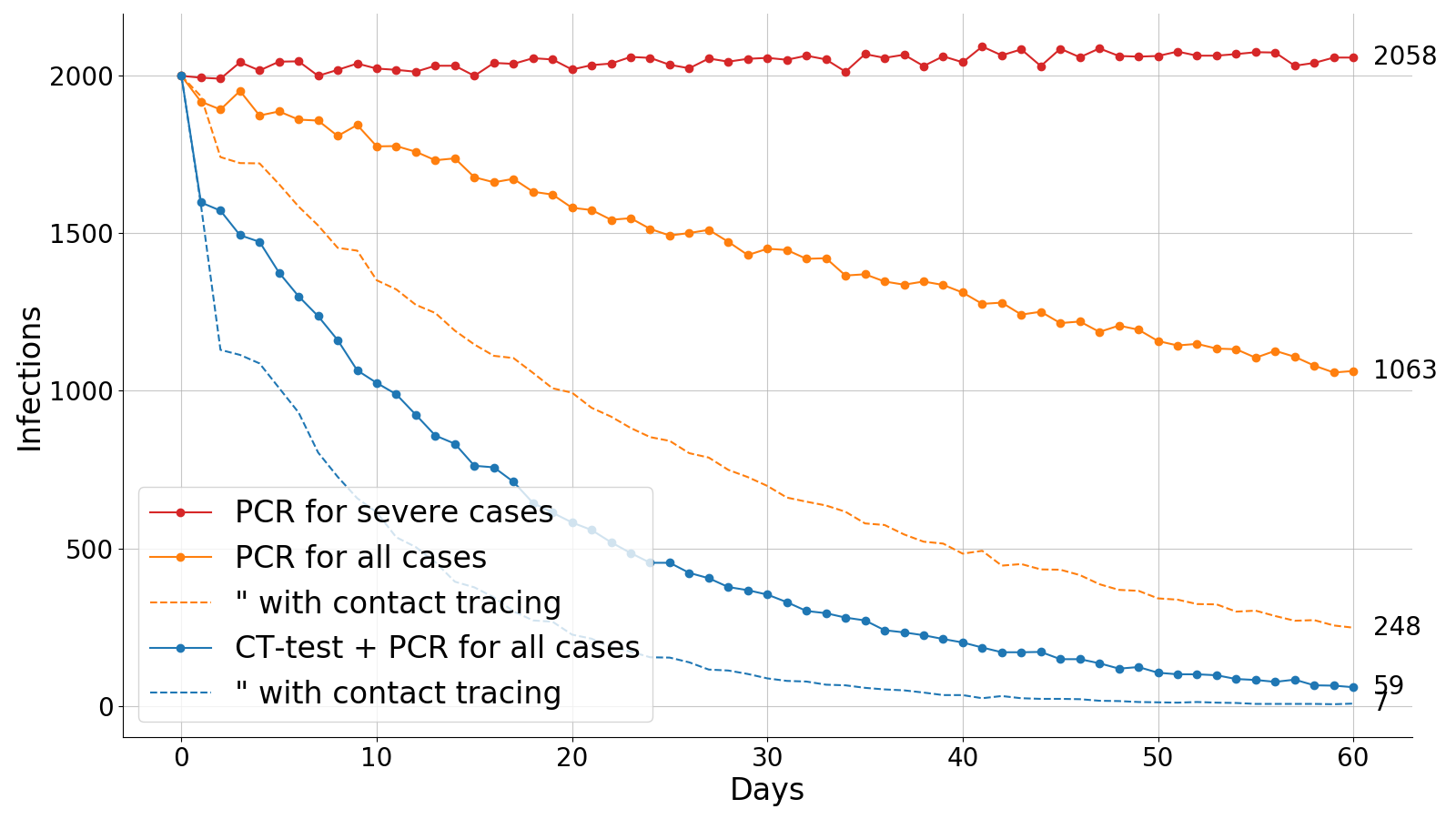}
\caption{As in Fig.1 but with a stronger social interventions so that $R^*=1.06$. We choose the $R^*$ so that in test strategy 1, infection rates are nearly constant. The first few days have sharp drop due to the change in testing and isolation strategy (see methods).}
\end{figure}

Results indicate that isolating based upon a positive CT-scan can address the false negative rate problem of PCR, accelerating the decline of cases to eradicate the disease. Confirmation of a COVID diagnosis (as opposed to another viral pneumonia) using a PCR test may take several days of testing. However, during this period a CT-scan with findings of viral pneumonia can already be used to effectively isolate positive cases. The number of positive test results expected is about 1 in 20 of those with early symptom tests, so the reduction in uncertainty is high and efforts to isolate are much lower than would otherwise be the case.

Using CT-scans leads to an infection reduction by $29\times$ and $35\times$ after 60 days compared to PCR for severe cases for moderate and strong social distancing respectively.  Compared to pervasive use of PCR the reduction is still large: $16\times$ and $18\times$. Contact tracing can be effective when tests are done for mild cases. Contact tracing reduces cases for PCR tests by $4\times$ for both, and for CT-scans by $12\times$ and $8\times$. Comparing PCR and CT-scan with contact tracing the reduction is $50\times$ and $35\times$. The ratios continue to diverge exponentially over time. Thus, rapid extinction of COVID is possible by combining CT-scans and contact tracing.

\begin{figure}[t]
\centering
\includegraphics[scale=0.24]{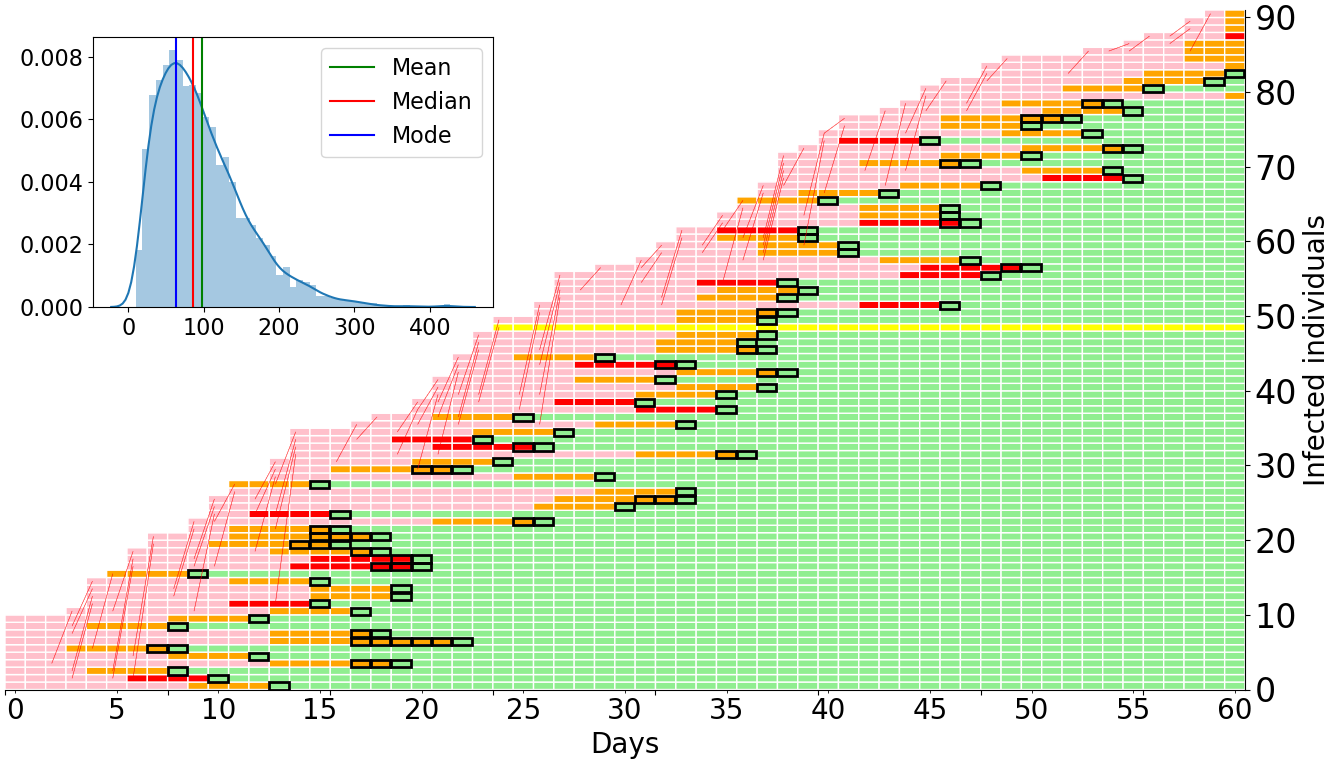}
 \vspace{-2ex}
\caption{
 Illustration of a simulation similar to Fig. 2, orange curve (widespread PCR tests, no CT use). Simulation starts at day 0 (horizontal axis) with 10 individuals and runs for 60 days with R* = 1.06. Vertical axis indicates each infected individual. Horizontal bars represent individual history from infection to symptoms, testing and isolation. Transmission shown by thin red line. Colors represent: pink $\rightarrow$ presymptomatic,  orange $\rightarrow$ mild, red $\rightarrow$ severe, yellow $\rightarrow$ asymptomatic, green $\rightarrow$ isolated, and black box $\rightarrow$ PCR test. Inset shows histogram of total number of infected individuals in the simulation (vertical axis is frequency), with the main panel showing an example with close to the median number of infections, 86.}
\end{figure}

\begin{figure}[t]
 \vspace{-2ex}
\centering
\includegraphics[scale=0.24]{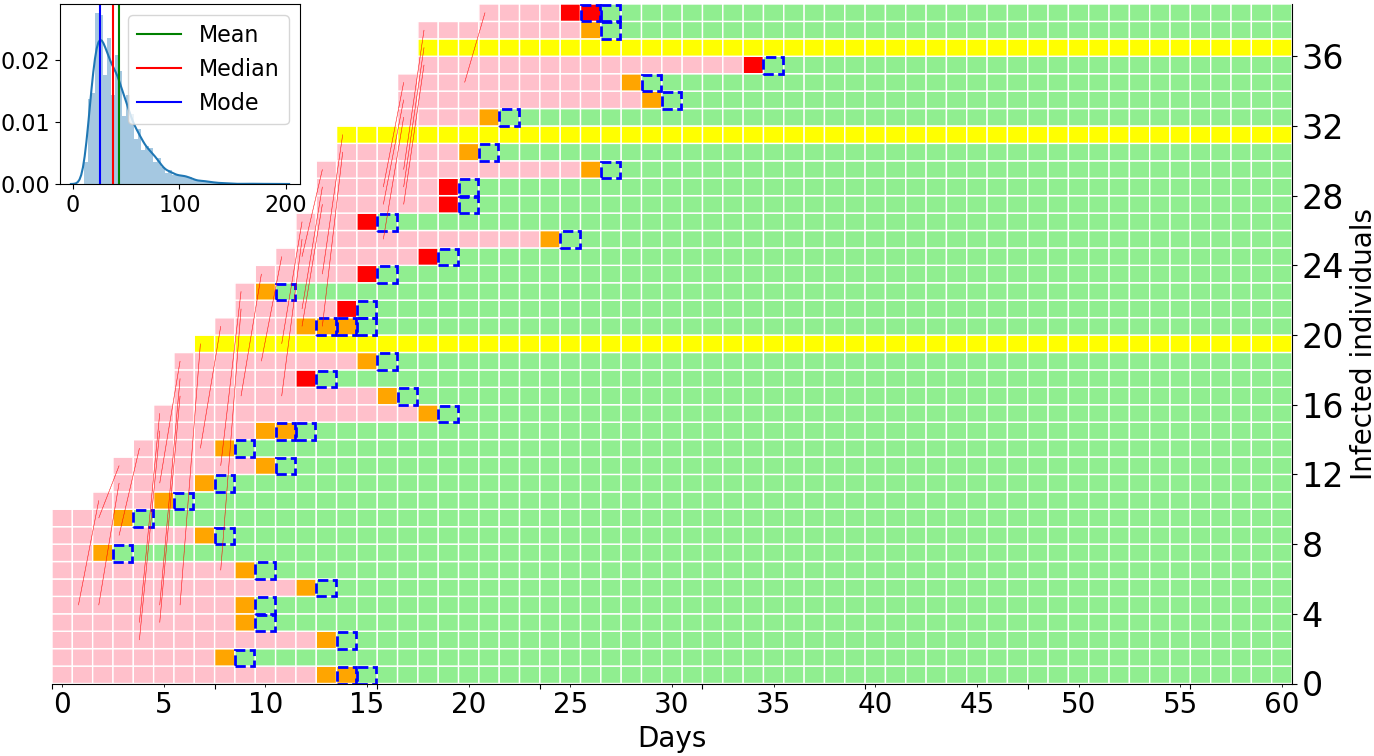}
 \vspace{-1ex}
\caption{ Illustration of a simulation similar to Fig. 2, blue curve (widespread CT-scan and PCR tests). Other parameters are the same as in Fig. 3. Dashed blue box $\rightarrow$ CT-scan, PCR for COVID diagnosis not shown. Fewer infected individuals (median 38) reflects the effectiveness of CT-scans and isolation in preventing transmission. Note that the last infection is on day 21, while with RT-PCR in Fig. 3 the outbreak doesn't stop in 60 days.}
 \vspace{-2ex}
\end{figure}

The five testing methods can also be characterized by effective reproduction ratios as follows: PCR for severe cases: $R^* \times 0.95 \rightarrow 5\%$ reduction, PCR for all cases: $R^* \times 0.90 \rightarrow 10\%$ reduction, PCR for all cases with contact tracing of 50\% of positive tested individuals: $R^* \times 0.80 \rightarrow 20\%$ reduction, CT-scans for all: $R^* \times 0.71 \rightarrow 29\%$ reduction, CT-scans for all with contact tracing of 50\% of positive tested individuals: $R^* \times 0.58 \rightarrow 42\%$ reduction. The difference between PCR by itself and CT-scans with PCR is $0.90 - 0.71 = 0.19$, and between both with $50\%$ contact tracing is $0.80 - 0.58 = 0.22$. $50\%$ contact tracing contributes a reduction of $0.10$ and $0.13$ for PCR and CT-scans respectively. 

\section*{Methods} 
Example simulations for the model are shown in Figures 3 and 4. The model is initialized with a number N of infected agents. Their incubation periods are determined individually by a Weibull distribution (Fig. 3) with $\alpha=2.04$ ($95\%$ CI: $1.80-2.32$) and $1/\lambda=0.103$ ($95\%$ CI: $0.096-0.111$)\cite{incubation}. The percentage of mild or moderate symptoms is $75\%$ and for severe or critical conditions $20\%$  \cite{joint report}. The persistent asymptomatic proportion continues to be debated (perhaps because of false negative PCR tests), here we set it to be $5\%$ for Figs. 1 and 2, and perform a sensitivity analysis by varying the percentage. The infectiousness of a patient as a function of time after infection is simulated by a beta distribution, $\beta(\tau) = B(\tau/\tilde{\tau})$ (Fig. 4), with parameters $\alpha_0=4$, $\beta_0=7$ and $\tilde{\tau}=15$ fit to established models \cite{transmission}. The reproduction rate, $R_t$, includes both the testing/isolation strategy and the effect of other social interventions, given by $R^*$ as indicated in the figures. For Figures 1 and 2 and the sensitivity analyses the model initially runs 20 days ($t = -20$ to $-1$) with test strategy 1 (described below) to represent under-testing in the initial phase. We continue ($t=0$ to $t=60$) with one of the following test and isolation strategies: 

\begin{enumerate}
    \item All severe and critical cases are tested. Mild and moderate cases are not tested. There is a 4-day delay between symptom onset and isolation due to hospital capacity and PCR test turnaround time. 
    \item All symptomatic individuals are tested. There is a 4-day delay between symptom onset and isolation for positive cases. Isolation does not occur for false negative cases, which remain infectious until a positive test result.
    \item Same as 2) with contact tracing: 50\% of contacts are traced and quarantined once a case is identified.
    \item All symptomatic cases are pre-screened by CT-scan, and isolated if positive while waiting for PCR to confirm. There is a 1-day delay between symptom onset and test/isolation.
    \item Same as 4) with contact tracing: 50\% of contacts are traced and quarantined once a case is identified.
\end{enumerate}

We adjust the initial number N of infections (day $-20$) so that so that for each $R^*$, 2,000 are infected at day $0$. All simulations compared with each other (same Figure) have the same $R^*$, and share the same simulation from $t=-20$ to $t=0$. Case numbers shown are the average of 60 runs. In the baseline scenario we conservatively allowed repeated tests to be performed every day, allowing PCR tests to progressively identify more positive cases. In one of the sensitivity analyses scenarios, mild cases are only tested once. Contact tracing is initiated when a case is identified by CT-scan or PCR, which happens at least one day after symptom onset, and more days after infection. The first day CT-scans are performed there is a backlog of cases that are waiting for PCR test results. All of these are scanned leading to a drop in transmission due to quarantines. 

\begin{figure}
\centering
\includegraphics[scale=0.32]{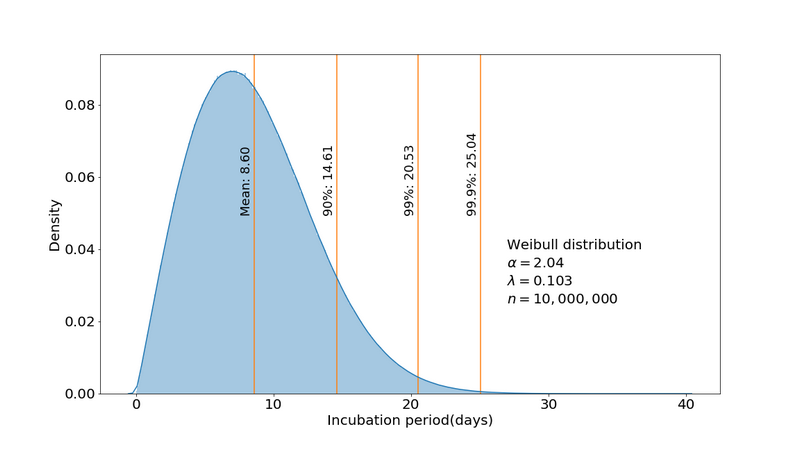}
 \vspace{-2ex}
\caption{The incubation period used in the model \cite{incubation}}
 \vspace{-2ex}
\end{figure}

\begin{figure}
\centering
\includegraphics[scale=2.50]{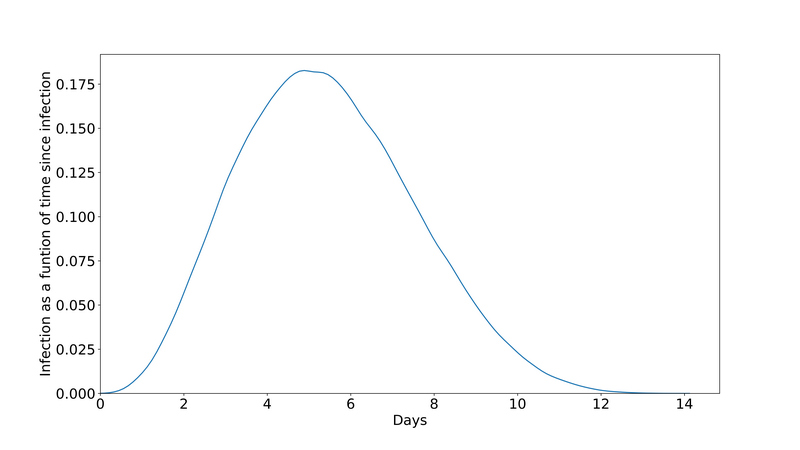}
 \vspace{-2ex}
\caption{The transmission dependence on time since infection, $\beta(\tau)$, used in the model \cite{transmission}. $\int d\tau \beta(\tau)=1$ and $R_t(\tau) = R^*\beta(\tau)$.}
\end{figure}

The percentage of false negative results of PCR and CT-scan are set to $30\%$ and $5\%$ respectively according to various reports \cite{CT, PCR-1,PCR-2}, and are varied in sensitivity analysis below. Isolated cases, whether in hospital settings or elsewhere, are assumed not to transmit (i.e. are not in home isolation). When contact tracing is enabled, every time a case is identified, either from PCR or CT-scan, a fraction of all the individuals infected by this individual agent are quarantined. The fraction is set to 0.5 in Fig. 1 and 2, and subject to sensitivity analysis below. Cases quarantined through contact tracing do not transmit.

We obtain changes in effective reproduction rate as follows: New cases at day $T$, can be represented as $N(T) = N R_{e}^{T/\tau_0}$, where $R_{e}$ is the effective reproduction rate, and $\tau_0 = 5$ is taken to be the reference case mean generation interval. Inverting and subtracting the reference case we have 
\begin{equation}
\delta R_{e} =  (N(T)/N)^{\tau_0/T} - (N^*(T)/N)^{\tau_0/T}
\end{equation}
where $N^*(T)$ is the reference simulation at day $T$. We have also directly calculated in the agent model the values of $R(t)$ including changes in the generation interval. Results are similar to Eq. 1, yielding an average of the two cases simulated: $R^* \times 0.95 \rightarrow 5\%$ reduction, PCR for all cases: $R^* \times 0.89 \rightarrow 11\%$ reduction, PCR for all cases with contact tracing of 50\% of positive tested individuals: $R^* \times 0.80 \rightarrow 20\%$ reduction, CT-scan for all: $R^* \times 0.72 \rightarrow 38\%$ reduction, CT-scan for all with contact tracing of 50\% of positive tested individuals: $R^* \times 0.61 \rightarrow 39\%$ reduction.

For additional confirmation we considered an analytic calculation that assumes one test for either severe or mild cases. The difference between pervasive PCR and CT-scan tests can then be calculated as:
\begin{equation} \begin{array}{ll}
(R_2 - R_4)/R^* &= (p_m+p_s)  \\
	& \times ( (F_p + (1-F_p) \sum_\tau w(\tau) \Xi(\tau + \Delta_p)) \\
	& -  (F_c + (1-F_c) \sum_\tau w(\tau) \Xi(\tau + \Delta_c)) )
\end{array} \end{equation}
where $w(\tau)$ is the Weibull distribution, $\Xi(\tau)$ is the cumulative distribution of $\beta(\tau)$. $\Delta_p = 4$, and $\Delta_c = 1$ are test result delays for PCR and CT-scans from symptom onset.  The result is $0.18$, in close agreement with simulations of reference and other scenarios in sensitivity analysis. 

\section*{Sensitivity analysis}

In sensitivity analysis we consider: Repeated versus single test, Asymptomatic proportion, CT-scan false negative rate, Contact tracing proportion, Stochasticity in the dynamics. 

\textbf{Without multiple testing.} In the baseline scenario PCR tests or CT-scans are performed repeatedly on successive days on false negative results to progressively identify more positive cases. If mild and moderate cases are tested only once, the results are as shown in Fig. 7 and 8. Conclusions are unaffected.

\textbf{Asymptomatic proportion.} For a sensitivity analysis we simulate 40\% mild/moderate cases, 10\% severe/critical cases and $50\%$ asymptomatic. Due to the comparatively low transmissibility of asymptomatic cases relative to symptomatic/presymtomatic cases \cite{transmission}, a similar observed transmission rate requires reducing the effectiveness of social distancing by increasing $R^*$ to $2.15$ and $1.83$. Results are shown in Figs. 9 and 10 and differ only weakly from the reference case in Figs. 1 and 2. Conclusions are unaffected. Note that we did not model the reported utility of CT-scan in detecting otherwise asymptomatic cases, which would provide an additional advantage for CT-scan in this scenario.

\begin{figure}
\centering
\includegraphics[scale=0.20]{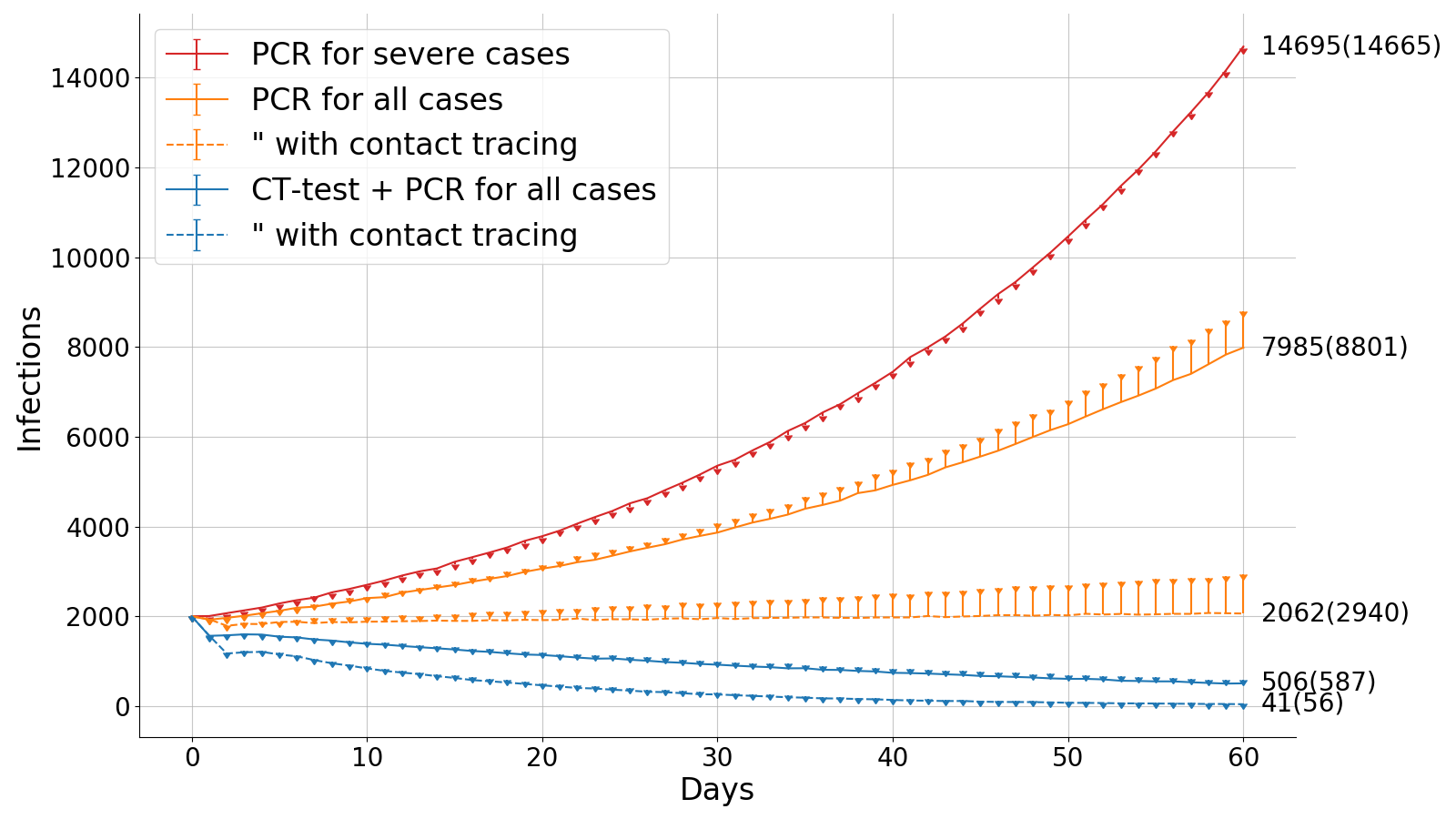}
 \vspace{-2ex}
\caption{As in Fig. 1, change to dots and numbers in parentheses show only one screening test for mild and moderate cases.}
\end{figure}
\begin{figure}
\centering
\includegraphics[scale=0.20]{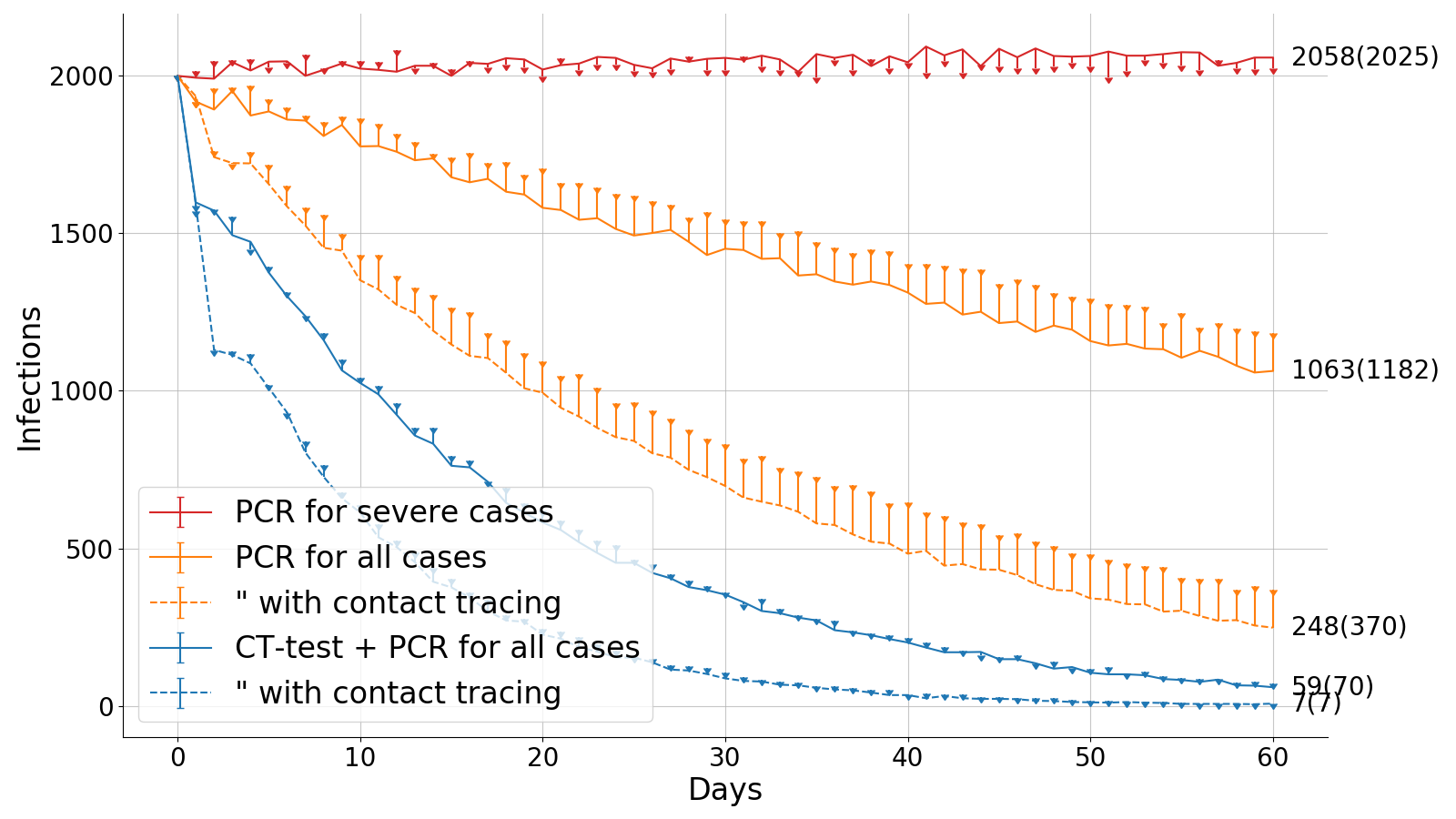}
 \vspace{-2ex}
\caption{As in Fig. 2, change to dots and numbers in parentheses show only one screening test for mild and moderate cases.}
\end{figure}

\textbf{CT-scan false negative rate.} We simulate scenarios where CT-scan has a higher false negative rate of $15\%$. Results are shown in Figs. 11 and 12. While the proportions are different in detail, the essential conclusions are unaffected. 

\begin{figure}
\centering
\includegraphics[scale=0.20]{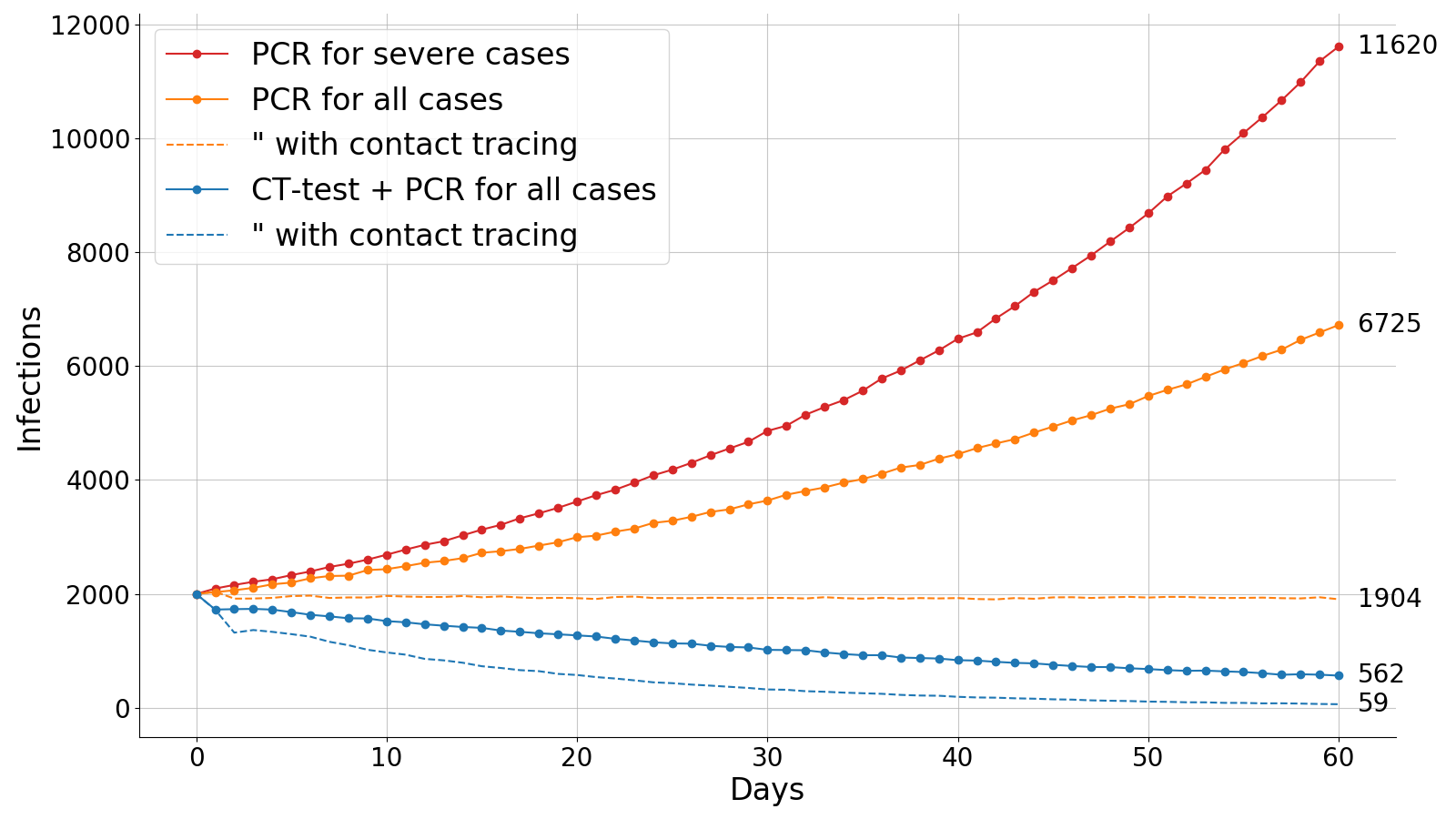}
 \vspace{-2ex}
\caption{Sensitivity analysis (change to dots, parentheses) for 40\% mild/moderate cases, 10\% severe/critical, and 50\% asymptomatic, with $R^*=2.12$ and initial infection numbers rescaled so that there are 2,000 infections on day 0.}
\end{figure}
\begin{figure}
\centering
\includegraphics[scale=0.20]{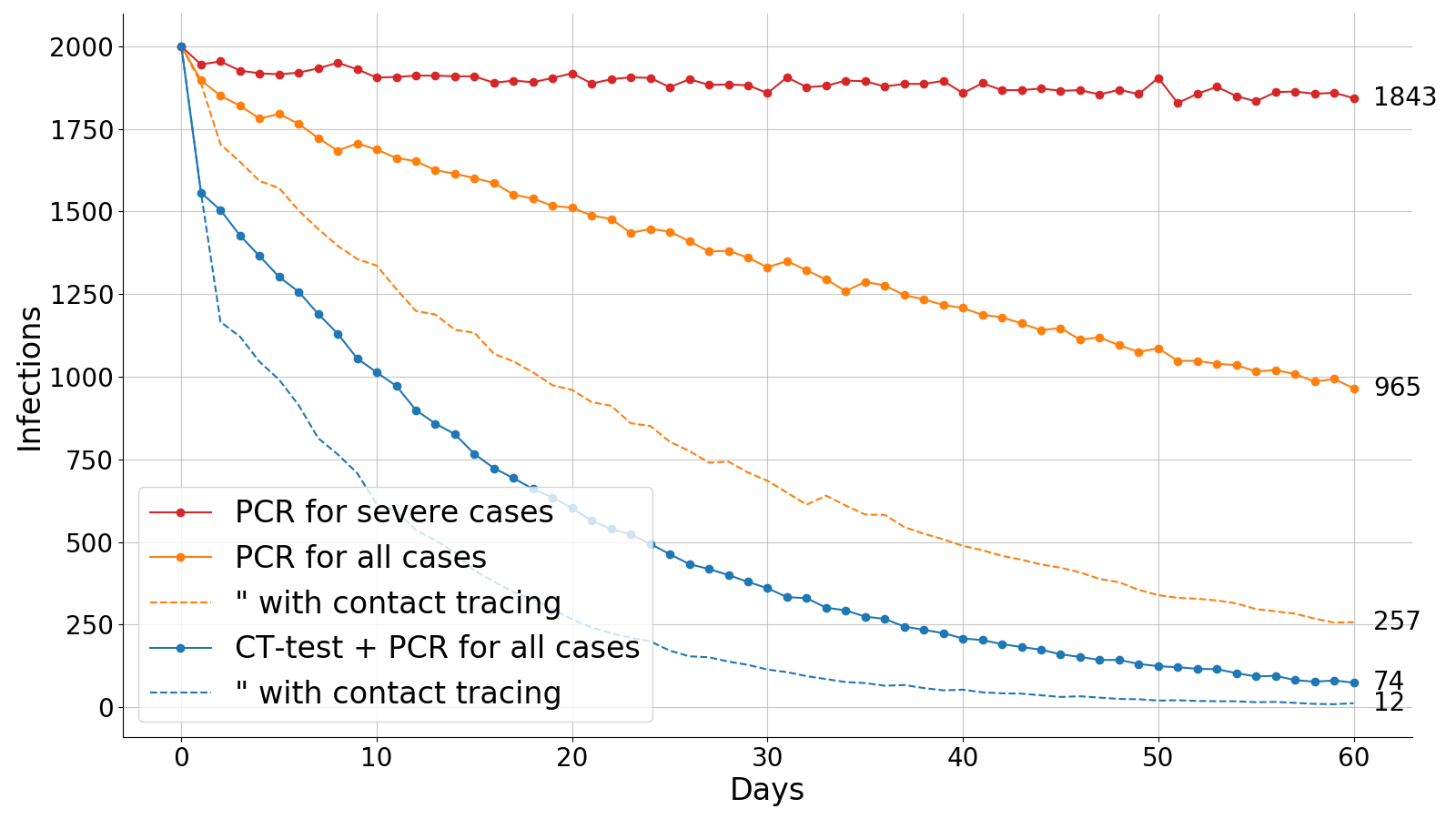}
 \vspace{-2ex}
\caption{Sensitivity analysis (change to dots, parentheses) for 40\% mild/moderate cases, 10\% severe/critical, and 50\% asymptomatic, with $R^*=1.83$ and initial infection numbers rescaled so that there are 2,000 infections on day 0.}
\end{figure}
\begin{figure}
\centering
\includegraphics[scale=0.20]{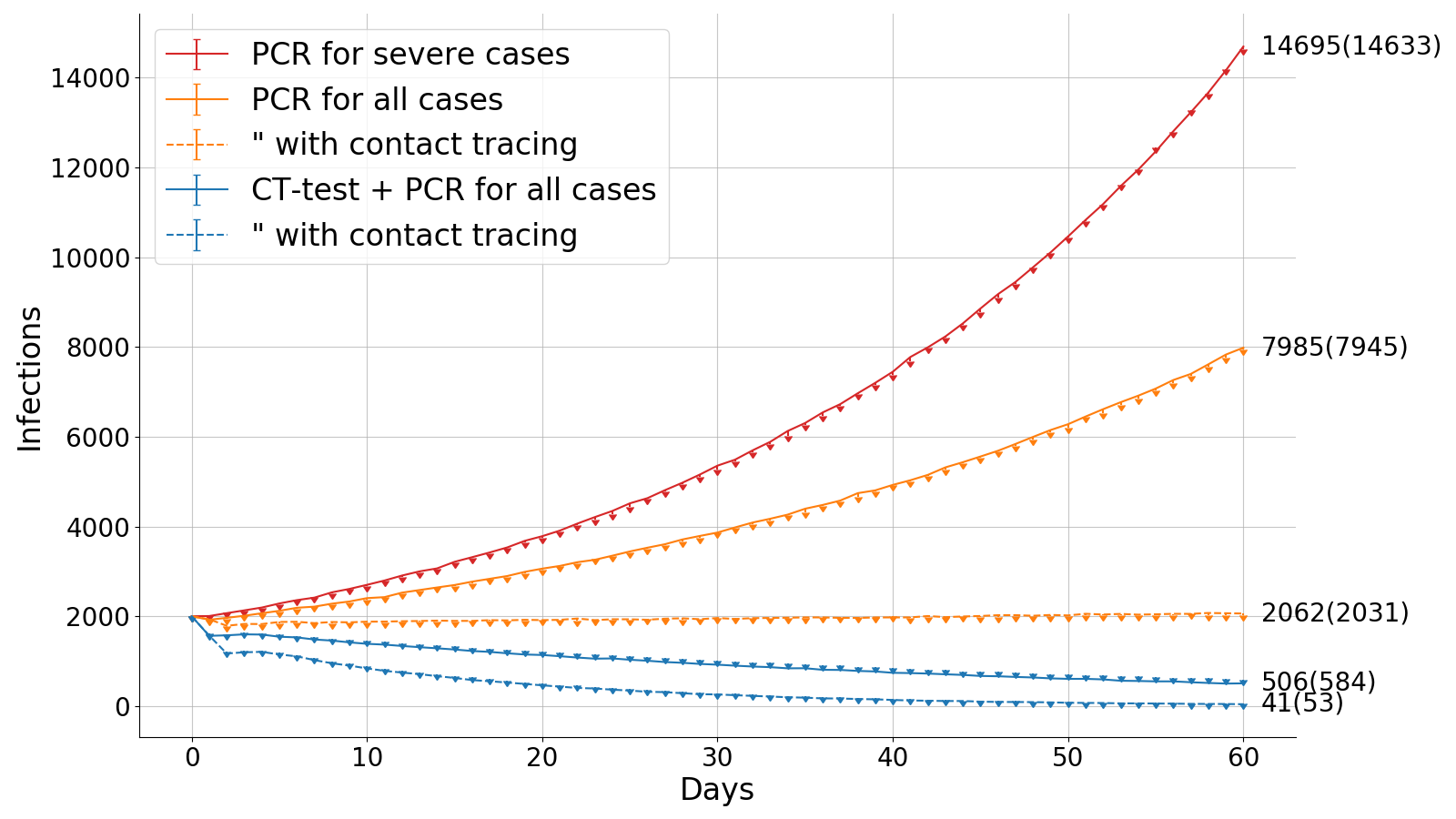}
 \vspace{-2ex}
\caption{Sensitivity analysis (change to dots, parentheses) for a higher CT-scan false negative rate of $15\%$. $R^*=1.25$.}
\end{figure}
\begin{figure}
\centering
\includegraphics[scale=0.20]{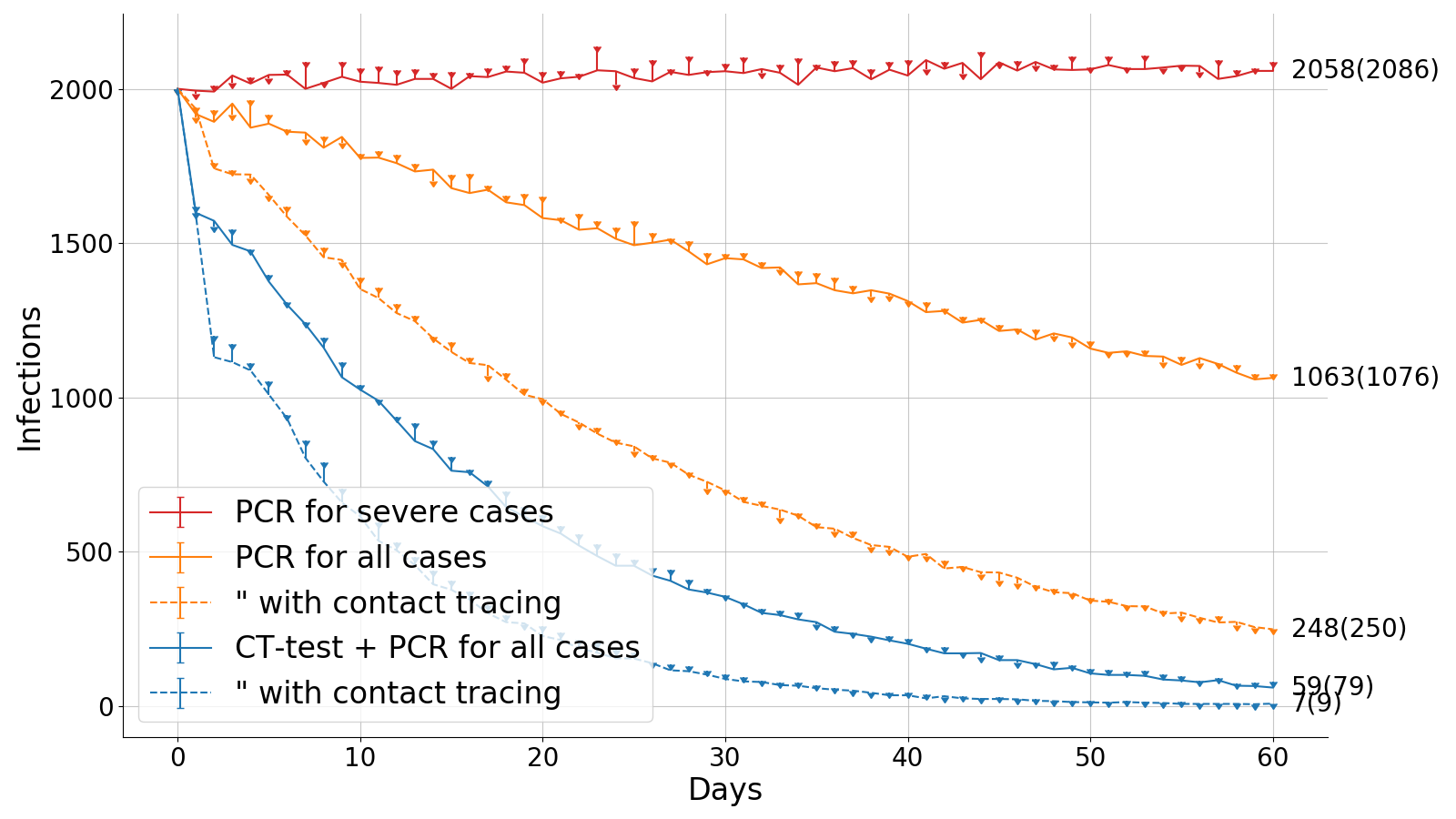}
 \vspace{-2ex}
\caption{Sensitivity analysis (change to dots, parentheses) for a higher CT-scan false negative rate of $15\%$. $R^*=1.06$.}
\end{figure}

\textbf{Contact tracing effectiveness.} Sensitivity analysis for the fraction of contacts traced. Results for $25\%$, $50\%$ and $75\%$ are shown in Figs. 13 and 14.  

\textbf{Stochasticity.} In order to give a sense of stochastic variation, we report the mean, $\bar{N}$, and standard deviation, $\sigma$, of runs for the reference cases. Simulations in Fig. 1, where $R^*=1.25$, have stochastic variation over 60 runs given by:
\begin{itemize}
    \item PCR for severe cases: $\bar{N}=14695$, $\sigma=600$
    \item PCR for all cases: $\bar{N}=7985$, $\sigma=400$
    \item PCR for all cases with contact tracing: $\bar{N}=2062$, $\sigma=160$
    \item CT + PCR for all cases: $\bar{N}$=506, $\sigma=50$
    \item CT + PCR for all cases with contact tracing: $\bar{N}=41$, $\sigma=15$
\end{itemize}

Simulations in Fig. 2, where $R^*=1.06$, have stochastic variation over 60 runs given by:
\begin{itemize}
    \item PCR for severe cases: $\bar{N}=2058$, $\sigma=260$
    \item PCR for all cases. $\bar{N}=1063$, $\sigma=168$
    \item PCR for all cases with contact tracing. $\bar{N}=248$, $\sigma=71$
    \item CT + PCR for all cases. $\bar{N}=59$, $\sigma=29$
    \item CT + PCR for all cases with contact tracing. $\bar{N}=7$, $\sigma=6$
\end{itemize}
Where standard deviations are comparable to the mean, multiple runs end with zero cases.

\textbf{Summary of Sensitivity Analysis.} Sensitivity analysis confirms the baseline case is representative. Our results imply that much more rapid extinction of COVID is possible by combining social distancing with CT-scans and contact tracing. 

\begin{figure}
\centering
\includegraphics[scale=0.20]{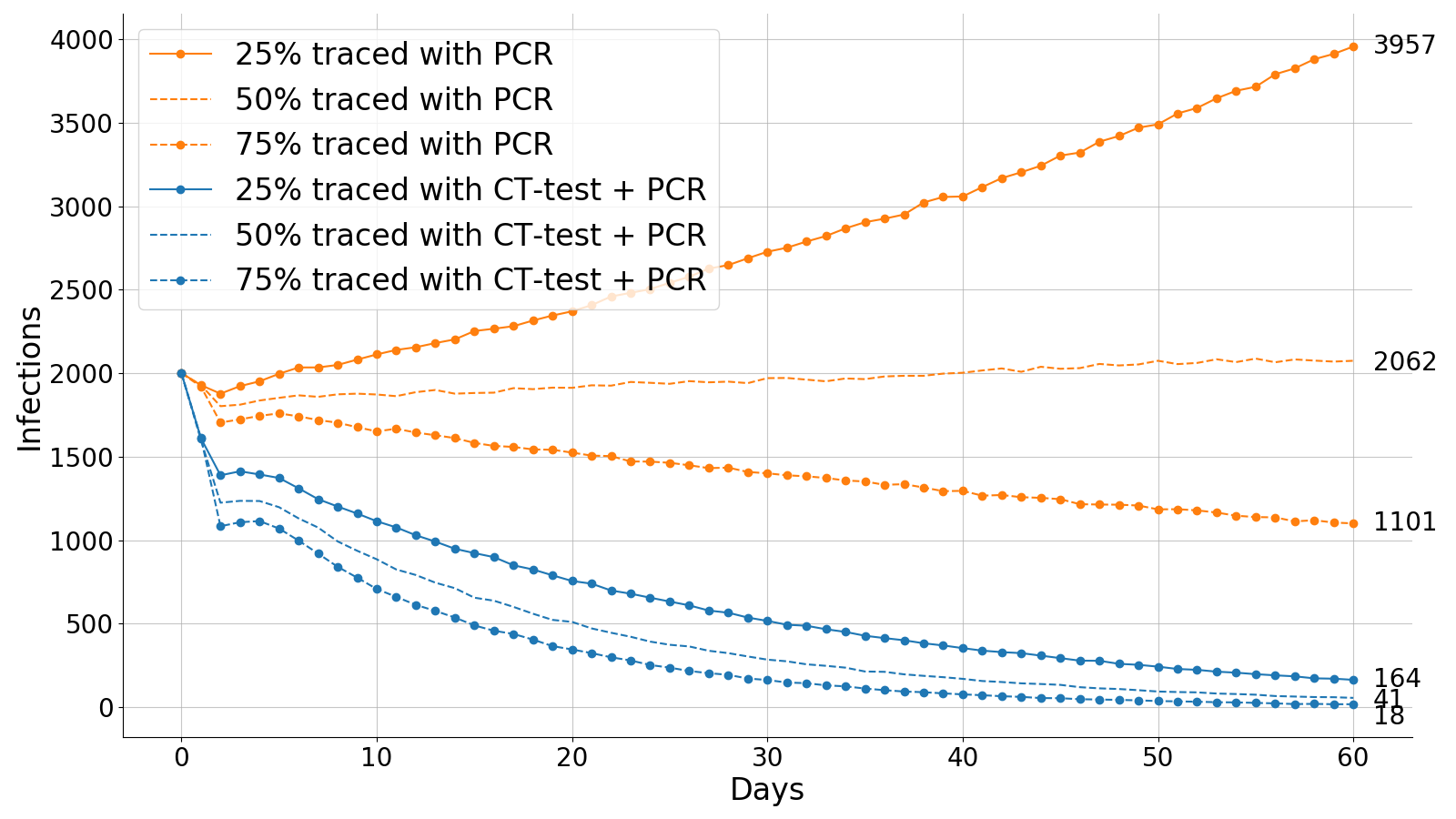}
 \vspace{-2ex}
\caption{Sensitivity analysis showing the effects of $25\%$, $50\%$ and $75\%$ of contacts traced and quarantined. $R^*=1.25$}
\end{figure}
\begin{figure}
\centering
\includegraphics[scale=0.20]{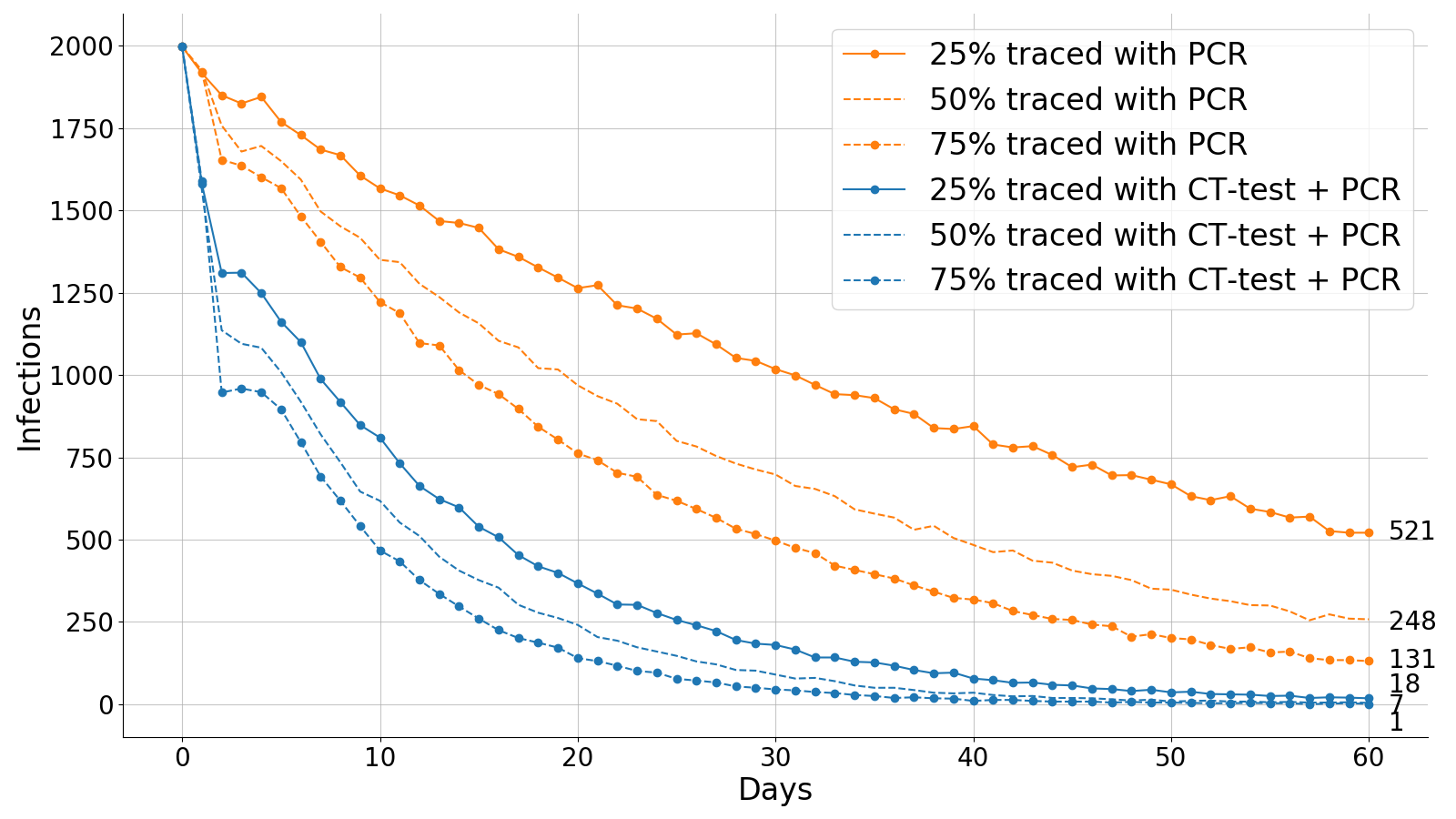}
 \vspace{-2ex}
\caption{Sensitivity analysis showing the effects of $25\%$, $50\%$ and $75\%$ of contacts are traced and quarantined, $R^*=1.06$.}
\end{figure}

\section*{Appendix: \label{App}}

\textbf{Cleaning and Decontamination.} Between CT-scans decontamination of surfaces and air exchange or decontamination is needed to avoid cross infection \cite{ACR}.  Guidelines recommend cleaning with disinfectant for over 30 or 60 min \cite{ACR,ChinaGuide} but can be performed within minutes by UV exposure \cite{UV}. Guidelines recommend 5 air exchanges \cite{airexchange}. HEPA purifiers are effective down to the size of viral particles \cite{HEPAfacts,size,size2}. Standard room size air purifiers perform 5 air exchanges per hour for, e.g., a 465 sq. ft. room \cite{honeywell}. An X min scan rate can be achieved with 60/X purifiers (10 min with 6 purifiers, 6 min with 10 purifiers). Exchange rates can be adjusted for smaller/larger rooms. Other mitigation practices are appropriate including masks, designated CT machines, peripheral devices at ambulatory providers, and mobile CT equipment.

\textbf{CT-scan FAQ.} For COVID-19 screening a low dose thin section CT-scan in supine position without contrast is appropriate. Typical pattern is unilateral, multifocal and peripherally-based ground glass opacities \cite{CT}. While concerns about costs are often raised, the cost of such a scan can be comparable to PCR. The potential for harms from radiation associated with a single LDCT is negligible with minimal risk of adverse consequences.  As with low dose CT for lung cancer screening, additional findings on the scans with diagnostic importance will be identified and can be managed according to well documented guidelines from the American College of Radiology.

\textbf{Acknowledgements:} We thank Guoping Sheng MD, Zeqing Xu MD, Zhiqiang Zhang MD, Rick Avila, Robert Senzig, Jenifer Siegelman MD MPH, and David Yankelevitz MD for helpful conversations.

%\end{multicols}

\end{document}